\documentclass[conference]{IEEEtran}
\usepackage{amsmath}
\usepackage{relsize}
\usepackage{framed}
\usepackage{graphicx}
\usepackage{caption}
\usepackage[flushleft]{threeparttable}
\usepackage{multirow}
\usepackage{hhline}
\usepackage{url}
\usepackage{rotating}
\usepackage{subcaption}
\usepackage{algpseudocode}
\usepackage{algorithm}
\usepackage{soul, color}
\soulregister\cite7
\soulregister\ref7

\newcommand{\R}{r}

\renewcommand{\hl}[1]{#1}

\begin{document}
\title{Parallel Pairwise Correlation Computation On Intel Xeon Phi Clusters}
\author{\IEEEauthorblockN{Yongchao Liu}
\IEEEauthorblockA{School of Computational \\Science \& Engineering\\
Georgia Institute of Technology\\
Atlanta, GA, USA\\
Email: yliu@cc.gatech.edu}
\and 
\IEEEauthorblockN{Tony Pan}
\IEEEauthorblockA{School of Computational \\Science \& Engineering\\
Georgia Institute of Technology\\
Atlanta, GA, USA\\
Email: tpan7@gatech.edu}
\and 
\IEEEauthorblockN{Srinivas Aluru}
\IEEEauthorblockA{School of Computational \\Science \& Engineering\\
Georgia Institute of Technology\\
Atlanta, GA, USA\\
Email: aluru@cc.gatech.edu}
}

% make the title area
\maketitle

\begin{abstract}
Co-expression network is a critical technique for the identification of inter-gene interactions, which usually relies on all-pairs correlation (or similar measure) computation between gene expression profiles across multiple samples. Pearson's correlation coefficient (PCC) is one widely used technique for gene co-expression network construction. However, all-pairs PCC computation is computationally demanding for large numbers of gene expression profiles, thus motivating our acceleration of its execution using high-performance computing. In this paper, we present LightPCC, the first parallel and distributed all-pairs PCC computation on Intel Xeon Phi (Phi) clusters. It achieves high speed by exploring the SIMD-instruction-level and thread-level parallelism within Phis as well as accelerator-level parallelism among multiple Phis. To facilitate balanced workload distribution, we have proposed a general framework for symmetric all-pairs computation by building bijective functions between job identifier and coordinate space for the first time. \hl{We have evaluated LightPCC and compared it to two CPU-based counterparts: a sequential C++ implementation in ALGLIB and an implementation based on a parallel general matrix-matrix multiplication routine in Intel Math Kernel Library (MKL) (all use double precision), using a set of gene expression datasets. Performance evaluation revealed that with one 5110P Phi and 16 Phis, LightPCC runs up to $20.6\times$ and $218.2\times$ faster than ALGLIB, and up to $6.8\times$ and $71.4\times$ faster than single-threaded MKL, respectively.} In addition, LightPCC demonstrated good parallel scalability in terms of number of Phis. Source code of LightPCC is publicly available at \url{http://lightpcc.sourceforge.net}.
\end{abstract}


% For peer review papers, you can put extra information on the cover
% page as needed:
% \ifCLASSOPTIONpeerreview
% \begin{center} \bfseries EDICS Category: 3-BBND \end{center}
% \fi
%
% For peerreview papers, this IEEEtran command inserts a page break and
% creates the second title. It will be ignored for other modes.
\IEEEpeerreviewmaketitle

\section{Introduction}
Co-expression networks have been frequently used to reverse engineer the whole-genome interactions between complex multicellular organisms by ascertaining common regulation and thus common functions. A gene co-expression network is represented as an undirected graph with nodes being genes and edges representing significant inter-gene interactions. Such a network can be constructed by computing linear (e.g. \cite{butte1999unsupervised}) or non-linear (e.g. \cite{margolin2006aracne} \cite{aluru2012reverse} \cite{lachmann2016aracne}) co-expression measures between paired gene expression profiles across multiple samples. As the first formal and wide-spread correlation measure \cite{lee1988thirteen} \cite{chandrashekar2014survey} %\cite{pratt1974correlation} \cite{brown1992survey} \cite{guyon2003introduction}  \cite{wang2013hybrid} \cite{chen2008characterization} \cite{d2005does}
, Pearson's correlation coefficient (PCC), or alias Pearson's $r$, is one widely used technique in co-expression network construction \cite{song2012comparison}. However, all-pairs PCC computation of gene expression profiles is not computationally trivial for genome-wide association study with large number of gene expression profiles across a large population of samples, especially when coupled with permutation tests for statistical inference. The importance of all-pairs PCC computation and its considerable computing demand motivated us to investigate its acceleration on parallel/high-performance computing architectures.

PCC statistically measures the strength of linear association between pairs of continuous random variables, but does not apply to non-linear relationship. Thus, we must ensure the linearity between paired data prior to the application of PCC. Given two random variables $u$ and $v$ of $l$ dimensions each, the PCC between them is defined as
\begin{equation}
\R(u, v) = \frac{\sum_{k=0}^{l-1} (u[k] - \bar{u})(v[k] - \bar{v})}{\sqrt{\sum_{k=0}^{l-1}(u[k] - \bar{u})^2 \sum_{k=0}^{l-1}(v[k]-\bar{v})^2}}
\label{equation:pearsonr}
\end{equation}
In Equation (\ref{equation:pearsonr}), $u[k]$ is the $k$-th element of $u$, while $\bar{u}$ is the mean of $u$ and equal to $\frac{1}{l}\sum_{k=0}^{l-1} u[k]$. Notations are likewise defined for $v$. Given a variable pair, the sequential implementation of Equation (\ref{equation:pearsonr}) has a linear time complexity proportional to $l$. Moreover, it is known that the absolute value of the nominator is always less than or equal to the denominator \cite{lee1988thirteen}. Thus, $\R(u, v)$ always varies in $[-1, +1]$. Concretely, \mbox{$\R(u, v) = 0$} indicates no linear relationship, \mbox{$> 0$} positive association and \mbox{$< 0$} negative association.

Although PCC is widely used in science and engineering, the acceleration of its computation using parallel/high-performance computing architectures has not yet been extensively investigated in the literature. Chang \textit{et al.} \cite{chang2009compute} used the CUDA-enabled GPU to accelerate all-pairs computation of PCC and computed pairwise PCC using the following standard reformulation:
\begin{equation}
\R(u, v) = \frac{\sum_{k=0}^{l-1} {u[k]\cdot v[k]} - l\cdot \bar{u}\cdot \bar{v}}{\sqrt{(\sum_{k=0}^{l-1} {u[k]^2} - l\cdot \bar{u}^2)( \sum_{k=0}^{l-1} {v[k]^2} - l\cdot \bar{v}^2})}
\end{equation}
This work was extended by \cite{kijsipongse2011efficient} to support GPU clusters, which adopted a master-slave model to manage workload distribution over multiple GPUs. Wang \textit{et al.} \cite{wang2013hybrid} adopted a hybrid CPU-GPU coprocessing model and reformulated each variable $u$ to a new representation $w$, which is defined as
\begin{equation}
w[k]= \frac{u[k] - \bar{u}}{|u[k] - \bar{u}|}
\label{equation:pearson_w}
\end{equation}
for each $k$ ($0\leq k < l$) in order to employ general matrix-matrix multiplication (GEMM) parallelization that has been well studied in parallel computing. Note that due to the commutative nature of pairwise PCC computation, this GEMM approach will cause a waste of half horsepower. Similarly, Wang \textit{et al.} \cite{wang2015full} also employed a parallel GEMM approach to accelerate the all-pairs computation of a given dataset $X$, but on a single Xeon Phi (Phi). Note that our approach is different from \cite{wang2015full}, because ours accelerates the overall computation over $X$ on a cluster of Phis.

In this paper, we present LightPCC, the first parallel and distributed algorithm to harness Phi clusters to accelerate all-pairs PCC computation. To achieve high speed, our algorithm explores instruction-level parallelism within SIMD vector processing units per Phi, thread-level parallelism over many cores per Phi, and accelerator-level parallelism across a cluster of Phis. Moreover, we have investigated a general framework for symmetric all-pairs computation to facilitate balanced workload distribution within and between processing elements (PEs), by pioneering to build a reversible and bijective relationship between job identifier and coordinate space in a job matrix. \hl{Using both artificial and real gene expression datasets, we have compared LightPCC to two CPU-based counterparts: a sequential C++ implementation in ALGLIB (http://www.alglib.net) and an implementation based on a parallel GEMM routine in Intel Math Kernel Library (MKL). Our experimental results showed that by using one 5110P Phi and 16 Phis, LightPCC is able to run up to $20.6\times$ and $218.2\times$ faster than ALGLIB, and up to $6.8\times$ and $71.4\times$ faster than singled-threaded MKL, respectively. In addition, LightPCC demonstrated good parallel scalability with respect to varied number of Phis.}
\section{Xeon Phi Architecture}
\label{sec:xeon_phi}
A Phi coprocessor is a many-core shared-memory computer \cite{jeffers2013intel}, which runs a specialized Linux operating system and provides full cache coherency over the entire chip. The Phi is comprised of a set of processor cores, and each core offers four-way simultaneous multithreading, i.e. 4 hardware threads per core. While offering scalar processing, each core also includes a newly-designed VPU which features a 512-bit wide SIMD instruction set architecture (ISA). Each vector register can be split to either 16 32-bit-wide lanes or 8 64-bit-wide lanes. The Phi does not provide support for legacy SIMD ISAs such as the SSE series. As for caching, each core locally has separate L1 instruction and data caches of size 32 KB each, and a 512 KB L2 cache. Moreover, all L2 caches across the entire chip are interconnected, through a bidirectional ring bus, to form a unified shared L2 cache of over 30 MB. In addition, there are two usage models for invoking Phis: offload model and native model. The offload model relies on compiler pragmas/directives to offload highly-parallel parts of an application to the Phi, while the native model treats a Phi as a symmetric multi-processing computer. As of today, Phis have been used to accelerate important computational problems  in diverse research fields such as bioinformatics \cite{liu2014swaphi} \cite{misra2015parallel} and machine learning \cite{jin2014training} \cite{viebke2015potential}. %Figure \ref{fig:xeon_phi} gives the architectural diagram of Phis.
%
% \begin{figure}[!h]
% \centering
% \includegraphics[width=\linewidth]{xeon_phi.eps}
% \caption{Architectural diagram of the Xeon Phi}
% \label{fig:xeon_phi}
% \end{figure}
% %
\section{Parallelized Implementation}
\subsection{Pearson's Correlation Coefficient Reformulation}
In the case of all-pairs computation, significantly more computation can be further reduced than pure pairwise computation. For instance, given a $l$-dimensional variable $u$, the values of $\sum_{k=0}^{l-1} (u[k] - \bar{u})$ and $\sqrt{\sum_{k=0}^{l-1}(u[k] - \bar{u})^2}$ could be repeatedly calculated up to $n-1$ times in the case of literal computing using Equation (\ref{equation:pearsonr}). Since these two values are only dependent on $u$, they can be computed once beforehand. We define $X=\{X_0, X_1, \dots, X_{n-1}\}$ to denote a set of $n$ $l$-dimensional variables and compute the new representation $U_i$ of $X_i$ as
\begin{equation}
	U_i[k] = \frac{X_i[k] - \bar{X_i}}{\sqrt{\sum_{k=0}^{l-1}(X_i[k] - \bar{X_i})}}
    \label{equation:u}
\end{equation}
In this way, the PCC between $X_i$ and $X_j$ is computed as
\begin{equation}
\R(U_i, U_j) = \sum_{k=0}^{l-1} U_i[k]\cdot U_j[k]
\label{equation:pearson2}
\end{equation}
From Equation (\ref{equation:pearson2}), we can see that if organizing all members of $U$ to form a $n\times l$ matrix $A$ with $U_i$ being row $i$ of $A$, we can realize the all-pairs computation over $U$ by multiplying matrix $A$ by its transpose, i.e. $R = A\times A^T$ via a GEMM algorithm. Note that because $R$ is symmetric, direct application of a GEMM algorithm will cause a waste of half compute power as noted before.

As mentioned above, Wang \textit{et al.} \cite{wang2013hybrid} also proposed a reformulation in order to benefit from parallel GEMM algorithms (refer to Equation (\ref{equation:pearson_w})). This reformulation computes the pairwise PCC between $X_i$ and $X_j$ as 
\begin{equation}
\R(U_i, U_j) = \frac{\sum_{k=0}^{l-1} U_i[k]\cdot U_j[k]}{\sqrt{\sum_{k=0}^{l-1} U_i[k]^2 \sum_{k=0}^{l-1}U_j[k]^2}}
\end{equation}
Using this equation, though a GEMM algorithm can be used to compute the nominator, the denominator has to be additionally computed.
\subsection{All-Pairs Computation Framework}
We consider the $n\times n$ job matrix to be a 2-dimensional coordinate space on the Cartesian plane, and define the left-top corner to be the origin, the horizontal $x$-axis (corresponding to columns) in left-to-right direction and the vertical $y$-axis (corresponding to rows) in top-to-bottom direction. 

\subsubsection{Non-symmetric all-pairs computation}
For non-symmetric all-pairs computation (non-commutative pairwise computation), the workload distribution over PEs (e.g. threads, processes, cores and etc.) would be relatively straightforward. 
This is because coordinates in the 2-dimensional matrix corresponds to distinct jobs. Specifically, given a coordinate $(y, x)$ ($0\leq x, y < n$), we can compute its unique job identifier $J_n(y, x)\in [0, n^2)$ as
\begin{equation}
	J_n(y, x) = yn +x
\end{equation}.
Reversely, given a job identifier $J_n(y, x)\in [0, n^2)$, we can compute its unique coordinate as
\begin{equation}
\begin{array}{ll}
x&	=J_n(y, x)\% n	\\
y&	=J_n(y, x) / n
\end{array}
\end{equation}
in the job matrix.
\subsubsection{Symmetric all-pairs computation}
Unlike non-symmetric all-pairs computation, it suffices by only computing the upper-triangle (or lower-triangle) of the job matrix for symmetric all-pairs computation (commutative pairwise computation). In this case, balanced workload distribution could be more complex than non-symmetric all-pairs computation. For workload distribution, one approach \cite{liu2009msa} is to allocate a separate job array, totally of $n(n+1)/2$ elements if the major diagonal is counted in, with element $i$ ($0\leq i < n(n+1)/2$) storing the coordinate of the $i$-th job and then let each PE access this array to obtain the coordinates of the jobs assigned to it. The major drawback of this approach is the extra memory consumed by the job array, since the memory overhead can be huge for large $n$. Another approach \cite{wang2013hybrid} is to use a policy designed for parallel matrix-matrix multiplication by discarding the redundant computing part, but could incur unbalanced workload distribution. In addition, some approaches (e.g. \cite{kijsipongse2011efficient}) use master-slave computing model.

In this paper, we propose a general framework for workload balancing in symmetric all-pairs computation. This framework works by assigning each job a unique identifier and then building a bijective relationship between a job identifier $J_n(y, x)$ and its corresponding coordinate $(y, x)$. This mapping is called {\tt direct bijective mapping} in our context. While facilitating balanced workload distribution, this mapping merely relies on bijective functions, which is a prominent feature distinguished from existing methods. To the best of our knowledge, in the literature bijective functions have not ever been proposed for workload balancing in symmetric all-pairs computation. In \cite{kiefer2010pairwise}, the authors used a very similar job numbering approach to ours in this study, but did not derive a bijective function for symmetric all-pairs computation. Our framework can be applied to cases with identical (e.g. our study) or varied workload per job (e.g. using a shared integer counter to realize dynamic workload distribution via remote memory access operations in MPI and Unified Parallel C (UPC) programming models) and is also particularly useful for parallel computing architectures with hardware schedulers such as GPUs and FPGAs. In the following, without loss of generality, we will interpret our framework relative to the upper triangle of the job matrix by counting in the major diagonal. Nonetheless, this framework can be easily adapted to the cases excluding the major diagonal.
\subsubsection{Direct bijective mapping}
\begin{figure}[!t]
\centering
\includegraphics[width=0.7\linewidth]{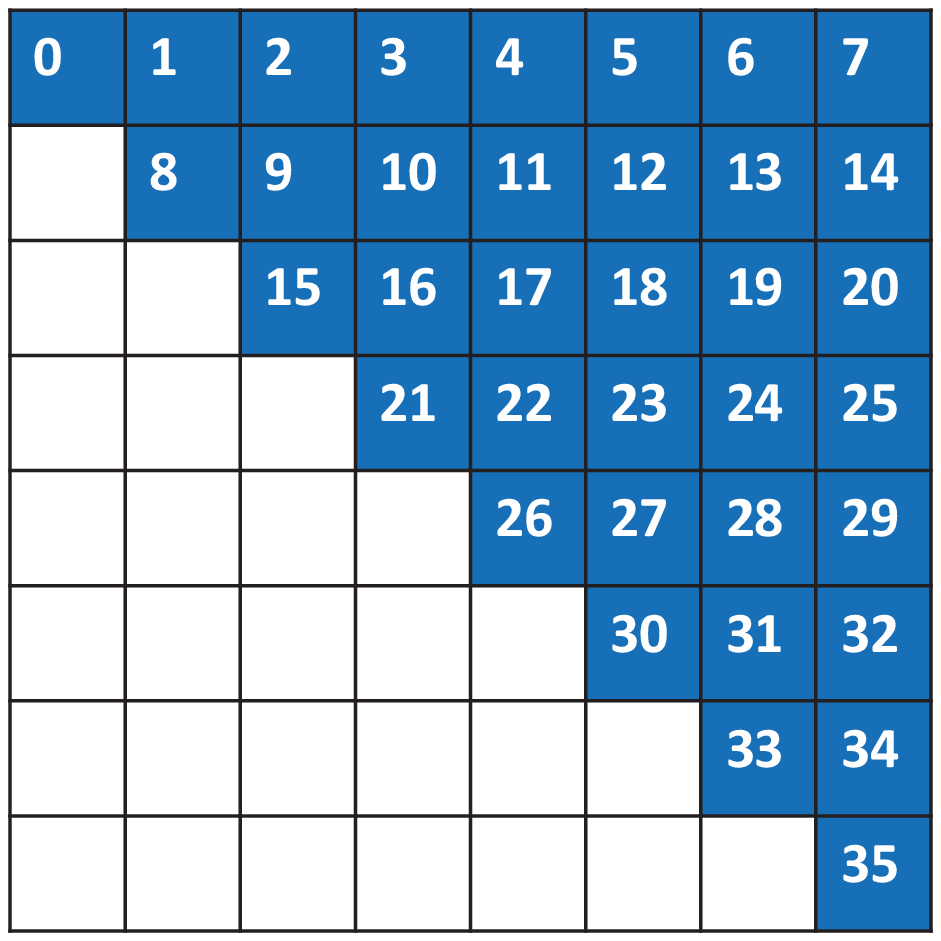}
\caption{An example direct bijective mapping between job identifier and coordinate space}
\label{fig:direct}
\end{figure}

Given a job $(y, x$) in the upper triangle, we compute its integer job identifier $J_n(y, x)$ as
\begin{equation}
\begin{array}{ll}
J_n(y, x) = F_n(y) + x - y,&0\leq y\leq x < n\\
\end{array}
\label{equation:j}
\end{equation}
for $n$ variables. In this equation, $F_n(y)$ is the total number of cells preceding row $y$ in the upper triangle and is computed as
\begin{equation}
F_n(y) = \frac{y(2n-y+1)}{2}
\label{equation:f}
\end{equation}
where $y$ varies in $[0, n]$ and there are two boundary cases needing to be paid attention to: one is when $y=0$ and the other is when $y=n$. When $y=0$, $F_n(0) = 0$ because no cell in the upper triangle appears before row $0$; and when $y=n$, $F_n(n) = n(n+1)/2$ because all cells in the upper triangle are included. In this way, we have defined Equation (\ref{equation:j}) based on our job numbering policy, i.e. all job identifiers vary in \mbox{$[0, n(n+1)/2)$} and jobs are sequentially numbered left-to-right and top-to-bottom in the upper triangle (see Fig. \ref{fig:direct} for an example).

Reversely, given a job identifier $J = J_n(y, x)$ ($0\leq J < n(n+1)/2$), we need to compute the coordinate $(y, x)$ in order to locate the corresponding variable pair. As per our definition, we have
\begin{equation}
\left\{
\begin{array}{l}
J \geq F_n(y) \Leftrightarrow y^2 - (2n+1)y + 2J \geq 0\\
J \leq F_n(y+1) - 1 \Leftrightarrow y^2 - (2n-1)y + 2(J+1) - 2n \leq 0\\
\end{array}
\right.
\label{equation:inequality_j}
\end{equation}
It needs to be stressed that there is surely an integer value $y$ satisfying these two inequalities based our job numbering policy mentioned above. By solving $J\geq F_n(y)$, we get
\begin{equation}
y \leq n + 0.5 - \sqrt{n^2 + n + 0.25 - 2J}
\label{equation:inequality_y_hi}
\end{equation}
This is because ($i$) $n^2 + n + 0.25 >2J$ and thus $J = F_n(y)$ has two distinct $y$ solutions theoretically, and ($ii$) all $0\leq y < n$ values are to the left of the symmetric axis $y = n + 0.5$, meaning strictly monotonically decreasing as a function of $y$. Meanwhile, by solving $J \leq F_n(y+1) - 1$, we get
\begin{equation}
y \geq n - 0.5 - \sqrt{n^2 + n + 0.25 - 2(J+1)}
\label{equation:inequality_y_low}
\end{equation}
Similarly, this is because ($i$) $n^2 + n + 0.25 > 2(J+1)$ and thus $J = F_n(y+1) - 1$ has two distinct $y$ solutions theoretically, and ($ii$) all $0\leq y < n$ values are to the left of the symmetric axis $y = n - 0.5$, meaning strictly monotonically decreasing as a function of $y$.

In this case, by defining $\Delta = \sqrt{n^2 + n + 0.25 - 2(J+1)}$, $\Delta' = \sqrt{n^2 + n + 0.25 - 2J}$ and $z = n - 0.5 - \sqrt{n^2 + n + 0.25 - 2(J+1)}$, we can reformulate Equations (\ref{equation:inequality_y_hi}) and (\ref{equation:inequality_y_low}) to be $z \leq y \leq z + 1 + \Delta - \Delta'$. In this case, because $\Delta < \Delta'$, we know that $[z, z + 1 + \Delta - \Delta']$ is a sub-range of $[z, z+1)$ and thereby have $z \leq y < z + 1$. As mentioned above, as a function of integer $y$, Equation (\ref{equation:inequality_j}) definitely has $y$ solutions as per our definition, meaning that at least one integer exists in 
$[z, z + 1 + \Delta - \Delta']$, which satisfies Equation (\ref{equation:inequality_j}). Meanwhile, it is known that there always exists one and only one integer in $[z, z+1)$ (can be easily proved) and this integer equals $\lceil z\rceil$, regardless of the value of $z$. Since $[z, z + 1 + \Delta - \Delta']$ is a sub-range of $[z, z + 1)$, we can conclude that Equation (\ref{equation:inequality_j}) has a unique solution $y$ that is computed as
\begin{equation}
	y = \lceil z \rceil = \Bigl\lceil n - 0.5 - \sqrt{n^2 + n + 0.25 - 2(J+1)} \Bigr\rceil
\label{equation:y}
\end{equation}
Having got $y$, we can compute the coordinate $x$ as
\begin{equation}
x = J + y - F_n(y)
\label{equation:x}
\end{equation}
based on Equation (\ref{equation:j}). Besides this theoretical proof, we also wrote a computer program to test its correctness.
%

% It needs to be stressed that in addition to the aforementioned theoretical proof, we also wrote a computer program to test its correctness using a set of $n$ values in practice. This program works as follows. For a specific $n$, given any coordinate $(y, x)$ in the upper triangle, we compute a job identifier $J$ using Equation (\ref{equation:j}). Because the job identifier corresponding to the coordinate $(y, x)$ is known beforehand (recalling our sequential job numbering), we compare the gold-standard job identifier with the computed $J$. If they are unequal, the bijective mapping is wrong and otherwise, we continue to the next step. Subsequently, we reversely compute a new coordinate $(y', x')$ from $J$ using Equations (\ref{equation:y}) and (\ref{equation:x}). If the coordinate $(y', x')$ is identical to its original $(y, x)$, the bijective mapping is correct and otherwise, wrong. Through our tests, the bijective functions work correctly.
%
\subsection{Tiled Computation on Xeon Phi}
Tiled computation is a frequently used technique in various applications accelerated by accelerators such as Cell/BEs \cite{sarje2013all}, GPUs \cite{liu2013cudasw++} and Phis \cite{liu2014swaphi}. This technique partitions a matrix into a non-overlapping set of equal-sized $t\times t$ tiles. In our case, we partition the job matrix and produce a tile matrix of size $m\times m$ tiles, where $m$ equals $\lceil n/t\rceil$. In this way, all jobs in the upper triangle of the job matrix are still fully covered by the upper triangle of the tile matrix. By treating a tile as a unit, we can assign a unique identifier to each tile in the upper triangle of the tile matrix and then build bijective functions between tile identifiers and tile coordinates in the tile matrix, similarly as we do for the job matrix.
\subsubsection{Computing tile coordinates}
As mentioned above, we have proposed a bijective mapping between job identifier and coordinate space. Because the tile matrix has an identical structure with the original job matrix, we can directly apply our aforementioned bijective mapping to the tile matrix. In this case, given a coordinate $(y_t, x_t)$ ($0\leq y_t \leq x_t < m$) in the upper triangle of the tile matrix, we can compute a unique tile identifier $J_m(y_t, x_t)$ as
\begin{equation}
\begin{array}{ll}
J_m(y_t, x_t) = F_m(y_t) + x_t - y_t,& 0\leq y_t \leq x_t < m\\
\end{array}
\label{equation:jt}
\end{equation}
where $F_m(y_t)$ is defined similar to Equation (\ref{equation:f}) as
\begin{equation}
F_m(y_t) = \frac{y_t(2m - y_t + 1)}{2}
\label{equation:ft}
\end{equation}
Likewise, given a tile identifier $J_t$ ($0\leq J_t < m(m+1)/2$), we can reversely compute its unique vertical coordinate $y_t$ as
\begin{equation}
y_t =  \Bigl\lceil m - 0.5 - \sqrt{m^2 + m + 0.25 - 2(J_t+1)}\Bigl\rceil
\label{equation:yt}
\end{equation}
and subsequently its unique horizontal coordinate $x_t$ as
\begin{equation}
x_t = J_t + y_t - F_m(y_t)
\label{equation:xt}
\end{equation}
\subsubsection{Multithreaded implementation}
Having got the coordinate $(y_t, x_t)$ of a tile, we can determine the coordinate range of all jobs per tile, relative to the original job matrix. More specifically, the vertical coordinate $y$ lies in $[y_t\times t, (y_t+1)\times t)$ and the horizontal coordinate $x$ in $[x_t\times t, (x_t+1)\times t)$. Consequently, the computation of a tile can be completed by looping over the coordinate ranges of both $y$ and $x$. Note that the jobs whose coordinate $y > x$ do not need to be computed since they lie beyond the upper triangle of the job matrix.
%We have given the pseudocode of a sequential tiled implementation in Algorithm \ref{alg:tiled}.
% %
% \begin{algorithm}
% \caption{Pseudocode of sequential tiled computation}
% \label{alg:tiled}
% \begin{algorithmic}[1]
% \fontsize{7}{7.5}\selectfont
% \Procedure{PearsonR}{$U, R, \dots$}
% \For{($J_t = 0; J_t < m(m+1)/2; ++J_t$)}

% 	\Comment{Compute the tile coordinate $(y_t, x_t)$}
%     \State {$y_t =  \Bigl\lceil m - 0.5 - \sqrt{m^2 + m + 0.25 - 2(J_t +1)}\Bigl\rceil$}
%     \State {$x_t = J_t + y_t - y_t(2m - y_t + 1)/2 $}
    
%     \Comment {Compute the tile}
%     \For{($y = y_t\times t; y < \min\{n, (y_t + 1)\times t\}; ++y$}
%     	\For{($x = x_t\times t; x < \min\{n, (x_t + 1)\times t\}; ++x$)}
%         	\If{$y \leq x$}
%             	\State {$r = 0$}
%             	\For{($k = 0; k < l; ++k$)}
%             		\State{$r += U_y[k]\cdot U_x[k]$}
%                 \EndFor
%                 \State{$R[y][x] = r$}
%             \EndIf
%         \EndFor
%     \EndFor
% \EndFor
% \EndProcedure
% \end{algorithmic}
% \end{algorithm}
% %

For all-pairs PCC computation, every job has the same amount of computation. In this case, the ideal load balancing policy is supposed to distributing identical number of jobs onto each PE. Herein, a thread on the Phi is referred to as a PE. From section \ref{sec:xeon_phi}, we know that each core on the Phi has four hardware threads and a two-level private L1/L2 cache hierarchy with caches being connected via a bidirectional ring bus to provide coherent caching on the entire chip. This non-uniform cache access reminds us that we should keep the active hardware threads per core sharing as much data as possible with the intention to improve caching performance. In these regards, our tiled computation schedules a tile to $t$ threads and lets these $t$ threads to compute the tile in parallel. Note that we must guarantee that the number of threads is a multiple of $t$ within the parallel region of our Phi kernel.

Algorithm \ref{alg:phi_tiled} shows the pseudocode of the Phi kernel of our tiled computation. Due to the limited amount of device memory, we are not able to entirely reside the resulting $n\times n$ correlation matrix $R$ in the Phi for large $n$ values. To address this problem, we partition the title identifier range $[0, m(m+1)/2)$ into a set of equal-sized non-overlapping sub-ranges and adopt a multi-pass execution model by letting one pass compute one sub-range, denoted by $[J_{start}, J_{end})$ for simplicity. To match this execution model, a result buffer $R'$ of size $(J_{end} - J_{start})\times t^2$ elements is allocated on the Phi and used to store the results of the $J_{end} - J_{start}$ tiles. For each tile, its $t^2$ results are consecutively placed in $R'$. Once a pass finishes, we transfer $R'$ to the host, then extract the results per tile and finally store them in the resulting matrix $R$ allocated on the host.

To benefit from the wide 512-bit SIMD vector instructions, we have aligned each $l$-dimensional variable in $U$ to 64-byte memory boundary on the Phi. In Algorithm \ref{alg:phi_tiled}, compiler directives are used to hint to the compiler to auto-vectorize the inner-most loop (lines 18 and 20 in Algorithm \ref{alg:phi_tiled}). Alternatively, we also manually vectorized the loop using SIMD instrinsic functions. The SIMD instrinsic functions used are \_mm512\_setzero\_ps/pd, \_mm512\_load\_ps/pd, \_mm512\_fmadd\_ps/pd, \_mm512\_mask3\_fmadd\_ps/pd, and \_mm512\_reduce\_add\_ps/pd
% \begin{itemize}
% \item \_mm512\_setzero\_ps/pd
% \item \_mm512\_load\_ps/pd
% \item \_mm512\_fmadd\_ps/pd
% \item \_mm512\_mask3\_fmadd\_ps/pd
% \item \_mm512\_reduce\_add\_ps/pd
% \end{itemize}
for single/double precision floating point. Interestingly, our manual-vectorization did not demonstrate obvious/significant performance advantage to auto-vectorization through our evaluations. More specifically, our manual-vectorization does run faster than auto-vectorization, but only by a tiny margin, on a single Phi. Considering that auto-vectorization is more portable than hard-coded SIMD intrinsic functions, we have used auto-vectorization all through our implementations.
\begin{algorithm}
\caption{Pseudocode of our tiled Xeon Phi kernel}
\label{alg:phi_tiled}
\begin{algorithmic}[1]
\fontsize{7}{7.5}\selectfont
\Procedure{mtPearsonR}{$U$, $R'$, $J_{start}$, $J_{end}$, $\dots$}
\State {\#pragma omp parallel}
\State{\{}
\State {$numGroups = omp\_get\_num\_threads() / t$}
\State {$tid = omp\_get\_thread\_num() \% t$}
\State {$gid = omp\_get\_thread\_num() / t$}
\For{($J_t = J_{start} + gid; J_t < J_{end}; J_t += numGroups$)}

	\Comment{Compute the tile coordinate $(y_t, x_t)$}
    \State {$y_t =  \Bigl\lceil m - 0.5 - \sqrt{m^2 + m + 0.25 - 2(J_t +1)}\Bigl\rceil$}
    \State {$x_t = J_t + y_t - y_t(2m - y_t + 1)/2$}
    
    \Comment{Compute offset $d$ in $R'$}
    \State {$d = (J_t - J_{start}) \times t^2 + tid$}
    
    \Comment{Compute its own $x$}
    \State {$x = x_t \times t + tid$}
    \If{($x < n$)}
    	\For {($y = y_t\times t; y < \min\{n, (y_t + 1)\times t\}; ++y$)}
           	\State {$r = 0$}
        	\If{($y \leq x$)}
    			\State{\#pragma vector aligned}
                \State{\#pragma simd reduction(+:r)}
                \For{($k = 0; k < l; ++k$)}
                	\State{$r += U_x[k] \cdot U_y]k]$}
                \EndFor
            \EndIf
            \State{$R'[d] = r$}
            \State{d += t}
        \EndFor
    \EndIf
\EndFor
\State {\}	//parallel region}
\EndProcedure
\end{algorithmic}
\end{algorithm}

\subsubsection{Asynchronous kernel execution}
As mentioned above, we rely on multiple passes of kernel execution to complete all-pairs computation. Conventionally, having completed one pass, we transfer the newly computed results to the host, and do not initiate a new kernel execution until having completed the processing of the new results. In this way, the co-processor will be kept idle, while we transfer and process the results on the host side. A better solution would be to employ asynchronous kernel execution, which enables concurrent execution of host-side tasks and device-side kernel execution. Fortunately, the offload model provides the {\tt signal} and {\tt wait} clauses to support for asynchronous data transfer and kernel execution. More specifically, the {\tt signal} clause enables asynchronous data transfer in {\tt \#pragma offload\_transfer} directives and asynchronous computation in {\tt \#pragma offload} directives. The {\tt wait} clause blocks the current execution until an asynchronous data transfer or computation has completed. Note that the {\tt signal} and {\tt wait} clauses are associated with each other via a unique value. In our implementation, we have used a double-buffering approach to facilitate asynchronous computation. Algorithm \ref{alg:async} gives the pseudocode of our asynchronous implementation.
\begin{algorithm}
\caption{Pseudocode of our asynchronous execution}
\label{alg:async}
\begin{algorithmic}[1]
\fontsize{7}{7.5}\selectfont
\Procedure{asyncKernelExecution}{}
\State{$J_{stop} = m(m+1)/2$}
\State{$J_{start} = 0$}
\State{$J_{end} = \min\{J_{stop}, J_{start} + maxNumTilesPerPass$\}}

\Comment{Initiate asynchronous kernel execution}
\State{\#pragma offload target(mic:id) signal (\&signalVar) $\cdots$}
	\State{\{$mtPearsonR(U, R'_{in}, J_{start}, J_{end})\}$}
    
    \Comment{Enter the core loop}
\While{(1)}

	\Comment{Wait for the kernel to complete and swap the buffers}
	\State{\#pragma offload target(mic:id) wait (\&signalVar) $\cdots$}
    \State{\{swap($R'_{in}, R'_{out}$)\}}
    \State{swap($R'_{in}, R'_{out}$)}
    
    \Comment{Save the previous range of tile identifiers}
    \State{$J'_{start} = J_{start}$}
    \State{$J'_{end} = J_{end}$}
    \If{($J_{end} \geq J_{stop}$)}
    	\State{break}
    \EndIf
    
    \Comment{Initiate asynchronous kernel execution}
    \State{$J_{start} += maxNumTilesPerPass$}
	\State{$J_{end} = \min\{J_{stop}, J_{start} + maxNumTilesPerPass\}$}
    
    \State{\#pragma offload target(mic:id) signal (\&signalVar) $\cdots$}
	\State{\{$mtPearsonR(U, R'_{in}, J_{start}, J_{end})$\}}
    
    \Comment{Process the results of the completed kernel}
    \State{$num = (J'_{end} - J'_{start}) \times t^2$}
    \State{Transfer $num$ elements in $R'_{out}$ from device to host}
    \State{Process the newly computed results on the host}
\EndWhile

\Comment{Process the results of the completed kernel}
\State{$num = (J'_{end} - J'_{start}) \times t^2$}
\If{($num > 0$)}
    \State{Transfer $num$ elements in $R'_{out}$ from device to host}
    \State{Process the newly computed results on the host}
\EndIf
\EndProcedure
\end{algorithmic}
\end{algorithm}
\subsection{Distributed Computing}
On Phi clusters, two distributed computing models can be used to develop parallel and distributed algorithms. One model is MPI offload model, which launches MPI processes just as an ordinary CPU cluster does. The difference is that one or more Phi coprocessors will be associated to a parental MPI process and this parental process will utilize offload pragmas/directives to interact with the affiliated Phis. In this model, communications between Phis have to be explicitly managed by their parental processes and it is not a necessity for Phis to be aware of the existence of remote communications between MPI processes. The other model is symmetric model, which treats a Phi as a regular computer interconnected to form a compute cluster. One advantage of symmetric model to MPI offload model is that symmetric model allows for the execution of existing MPI programs designed for CPU clusters to be directly executed on Phi clusters, with no need of re-programming the code. Nonetheless, considering different architectural features between CPUs and Phis, some amount of efforts may have to be devoted to performance tuning on Phi clusters.

In LightPCC, we used MPI offload model with the requirement of one-to-one correspondence between MPI processes and Phis. This pairing is straightforward for the cases launching one MPI process into one node. However, it would become more complex when a node has multiple Phis available and multiple processes running. This is because multiple processes launched into the same node have no idea about which Phi should be associated to themselves. To address this problem, we have used the registration-based management mechanism used by \cite{liu2014swaphi} for paring MPI processing and Phis.

Our distributed implementation is also based on tiled computation on the Phi. Given $p$ MPI processes, we evenly distribute tiles onto the $p$ processes with the $i$-th ($0\leq i < p$) process assigned to compute the tiles whose identifiers are in $[i\times \lceil m(m+1)/2p\rceil, (i + 1)\times \lceil m(m+1)/2p\rceil)$. Within each process, we adopt the same asynchronous control workflow with the computation for single Phis (see Algorithm \ref{alg:async}) and execute the same tiled computation kernel on the affiliated Phi in each pass (see Algorithm \ref{alg:phi_tiled}). Note that the initialization of variables $J_{start}$ and $J_{stop}$ (lines 2$\sim$3 in Algorithm \ref{alg:async}) must be changed accordingly for each process. Concretely, $p_i$ should initialize $J_{start}$ to be $i\times \lceil m(m+1)/2p\rceil$, and $J_{stop}$ to be $(i + 1)\times \lceil m(m+1)/2p\rceil$.
\subsection{Variable Transformation on Xeon Phi}
As mentioned above, we reformulate the computation of PCC by transforming each original variable $X_i$ to a new representation $U_i$ based on Equation (\ref{equation:u}). This variable transformation for input set $X$ only needs to be done once beforehand, and is also embarrassingly parallel since the transformation per variable is mutually independent. On the other hand, each variable requires identical amount of computation since these variables have the same dimension. In these regards, we parallelize the variable transformation by evenly distributing variables onto all threads on the Phi.

Algorithm \ref{alg:var_transform} gives the pseudocode of the Phi kernel of our variable transformation. In Algorithm \ref{alg:var_transform}, for each variable $X_i$ the transformation consists of three steps. Step 1 (lines 7$\sim$13) computes the mean of all elements in $X_i$ and requires $l$ unit arithmetic operations. Step 2 (lines 14$\sim$20) computes the variance of all elements and takes $2l$ unit arithmetic operations if considering a fused multiply-add operation as a unit one. Step 3 (lines 21$\sim$25) finishes the transformation of $X_i$ to $U_i$ and also needs $2l$ unit arithmetic operations. Therefore, the total computational cost of variable transformation can be estimated as $5l$ unit arithmetic operations. On the other hand, for symmetric all-pairs computation using Equation (\ref{equation:pearson2}), its computational cost can be estimated to be $ln(n+1)/2$ unit arithmetic operations. Consequently, the overall computational cost of our method can be estimated to be $5ln + ln(n+1)/2$ unit arithmetic operations.
\begin{algorithm}
\caption{Pseudocode of our variable transformation kernel}
\label{alg:var_transform}
\begin{algorithmic}[1]
\fontsize{7}{7.5}\selectfont
\Procedure{variableTransformation}{($X, U$)}

\State{\#pragma omp parallel}
\State{\{}
	\State{$tid = omp\_get\_thread\_num()$}
    \State{$chunk = \lceil \frac{n}{omp\_get\_num\_threads()}\rceil$}
	\For{($ i = tid\times chunk; i < \min\{n, (tid + 1)\times chunk \}; ++i$)}
    
    	\Comment{Compute the mean}
    	\State{$mean = 0$}
        \State{\#pragma vector aligned}
        \State{\#pragam simd reduction(+:mean)}
        \For{($k = 0; k < l; ++k$)}
        	\State{$mean += X_i[k]$}
        \EndFor
        \State{$mean /= l$}
        
        \Comment{Compute the variance}
        \State{$variance = 0$}
        \State{\#pragma vector aligned}
        \State{\#pragam simd reduction(+:variance)}
        \For{($k = 0; k < l; ++k$)}
        	\State{$variance += (X_i[k] - mean)^2$}
        \EndFor
        \State{$variance = \frac{1.0}{\sqrt{variance}}$}
        
        \Comment{Compute $U_i$ (in place)}
      	\State{\#pragma vector aligned}
        \State{\#pragam simd}
        \For{($k = 0; k < l; ++k$)}
        	\State{$U_i[k] = (X_i[k] - mean)\times variance$}
        \EndFor
        
	\EndFor
\State{\}//parallel region}
\EndProcedure
\end{algorithmic}
\end{algorithm}
\section{Performance Evaluation}
We evaluated LightPCC from \hl{three perspectives: ($i$) performance comparison with the sequential ALGLIB (version 3.10.0), ($ii$) performance comparison with an implementation based on the {\tt cblas\_dgemm} GEMM routine in MKL, which first transforms variables based on Equation (\ref{equation:u}) (refer to Algorithm \ref{alg:var_transform}) and then applies GEMM based on Equation (\ref{equation:pearson2}), and ($iii$)} parallel scalability evaluation on a Phi cluster, using a set of artificial and real gene expression datasets. All tests are conducted on 8 compute nodes in CyEnce HPC Cluster (Iowa State University), where each node has two Intel E5-2650 8-core 2.0 GHz CPUs, two 5110P Phis (each has 60 cores and 8 GB memory) and 128 GB memory. \hl{Every program is} compiled by Intel C++ compiler v15.0.1 with option {\tt -fast} enabled. Meanwhile, when two processes run in a node, we used the environment variable {\tt I\_MPI\_PIN\_PROCESSOR\_LIST} to guide Intel MPI runtime system to pin two processes per node to distinct CPUs (recall that a node has two CPUs).

For LightPCC, we set the tile size to $4\times 4$ (i.e. $t=4$) and configure each core to run four hardware threads associated with the {\tt compact} OpenMP thread affinity mode (i.e. 236 actives threads on 59 cores). In this implementation, we schedule one tile to a core at a time, and let all four threads per core compute the same tile in parallel, with one thread processing one column. In this way, the four threads per core will access the same row variable, thereby improving data sharing. However, even though a 5110P Phi executes four hardware threads per core in order \cite{jeffers2013intel}, the four threads on each core are actually scheduled individually and independently by the operating system. Hence, we cannot guarantee that the four hardware threads per core always work on the same tile at any instant. In this regard, we have introduced a software-based centralized barrier \cite{mellor1991algorithms}, which is implemented using the atomic intrinsic function {\tt \_\_sync\_fetch\_and\_sub}, in order to synchronize the four hardware threads per core each time they finish their computation on a tile (note that cores are configured to have independent software barriers). Unfortunately, we observed slight performance decrease after using software barriers through our evaluations. In this regard, we have decided not to use software barriers both in our implementation and following tests. In addition, \hl{for each evaluated program, we used double-precision floating point for fair comparison and averaged its five runs to get the runtime.}
\subsection{Evaluation on Artificial Gene Expression Data}
We first evaluated the performance of LightPCC, ALGLIB \hl{and the implementation using MKL (refer to as MKL for short in the following)} using three artificial gene expression datasets by randomly generating gene expression values in $[0, 1]$. This is reasonable because the runtime of PCC computation is merely subject to $n$ and $l$ and independent of specific values. They are randomly generated by setting $n$ to 16,000 (16K), 32,000 (32K) or 64,000 (64K) and $l$ to 5,000 (5K).

Table \ref{tab:artificial_alglib} shows the performance comparison between LightPCC and ALGLIB. \hl{Compared to ALGLIB, LightPCC runs $12.9\times$, $17.4\times$ and $20.6\times$ faster using one Phi and $160.1\times$, $190.2\times$, and $218.2\times$ faster using 16 Phis for $n$=16K, $n$=32K and $n$=64K, respectively. Moreover, the speedup gradually increases as $n$ grows larger.} \hl{It is worth mentioning} that many applications require determining the statistical significance of pairwise correlation. For this purpose, permutation test is a frequently used approach for statistical inference. However, this approach needs to repeatedly permute vector variables at random and compute pairwise correlation from the random data, where the more iterations (typically \mbox{$\geq$1,000} iterations) are conducted, the more precise statistical results (e.g. $P$-value) can be expected (except for the cases of complete permutation tests that rarely happen). In this case, the runtime with a specified number of permutation tests can be roughly inferred from the runtime per iteration and the number of iterations conducted.
\begin{table}[!t]
\centering
\caption{Comparison with ALGLIB on artificial data}
\label{tab:artificial_alglib}
\begin{tabular}{|p{1.1cm}||p{0.6cm}||p{0.7cm}||p{0.7cm}||p{0.6cm}||p{0.6cm}||p{0.6cm}|}
\hline
\multirow{2}{*}{Program}& \multicolumn{3}{|c||}{Time (s)}&	\multicolumn{3}{|c|}{Speedup}\\
\hhline{~------}
&16K&	32K&	64K&	16K&	32K&	64K\\
\hline
ALGLIB&	355.0 & 1,451.1 & 5,891.6& $-$&	$-$&	$-$\\
\hline
1 Phi&	27.4 & 83.2 & 285.4& 	12.9  & 17.4  & 20.6\\
\hline
2 Phis&	15.6 & 55.5 & 203.7&	22.8  & 26.2  & 28.9 \\
\hline
4 Phis&	7.8  & 29.6 & 103.8&	45.7  & 49.1  & 56.8\\
\hline
8 Phis&	3.9  & 15.1 & 52.4&		90.2  & 96.1  & 112.5 \\
\hline
16 Phis&2.2  & 7.6  & 27.0&		160.1 & 190.2 & 218.2\\
\hline
\end{tabular}
\end{table}

\hl{Table \ref{tab:artificial_mkl} compares LightPCC with MKL. Compared to single-threaded MKL, LightPCC runs up to $6.8\times$ faster on one Phi and up to $71.4\times$ faster on 16 Phis. Compared to 16-threaded MKL, LightPCC yields inferior performance on $<4$ Phis but superior performance when using $\ge 4$ Phis, where the maximum speedup 5.3 is reached in the case applying 16 Phis to the 64K dataset. Moreover, for both comparisons, the speedups of LightPCC over MKL are observed to increase as $n$ becomes larger. In addition, we evaluated MKL on a single Phi by only using the smallest 16K dataset, because a Phi does not own enough device memory to have the correlation matrix $R$ reside entirely in memory. This is also one reason why we adopted the aforementioned multi-pass solution with asynchronous kernel execution. Through our test, the Phi-based MKL took 8.8 seconds to finish the computation, outperforming our LightPCC by a factor of 3.11 on a single Phi. Nevertheless, it should be noted that our algorithm is designed to enable fast processing of very large datasets by overcoming single Phi device memory limitation and leveraging Phi clusters. In contrast, these features cannot be easily realized by MKL.}
\begin{table}[!t]
\centering
\caption{Speedups over Intel MKL on artificial data}
\label{tab:artificial_mkl}
\begin{tabular}{|p{1.1cm}||p{0.6cm}||p{0.7cm}||p{0.7cm}||p{0.6cm}||p{0.6cm}||p{0.6cm}|}
\hline
\multirow{2}{*}{Program}& \multicolumn{3}{|c||}{MKL (1 core)}&	\multicolumn{3}{|c|}{MKL (16 cores)}\\
\hhline{~------}
&16K&	32K&	64K&	16K&	32K&	64K\\
\hline
\hline
1 Phi&	4.3  & 5.7  & 6.8  & 0.3 & 0.4 & 0.5\\
\hline
2 Phis&	 7.7  & 8.6  & 9.5  & 0.6 & 0.6 & 0.7 \\
\hline
4 Phis&	15.4 & 16.2 & 18.6 & 1.2 & 1.2 & 1.4 \\
\hline
8 Phis&	30.3 & 31.7 & 36.8 & 2.3 & 2.4 & 2.7\\
\hline
16 Phis&	53.8 & 62.7 & 71.4 & 4.2 & 4.7 & 5.3\\
\hline
\end{tabular}
\end{table}

\begin{figure}[!t]
\centering
\begin{subfigure}{0.49\linewidth}
\fontsize{7}{7.1}\selectfont
\includegraphics[width=0.985\linewidth]{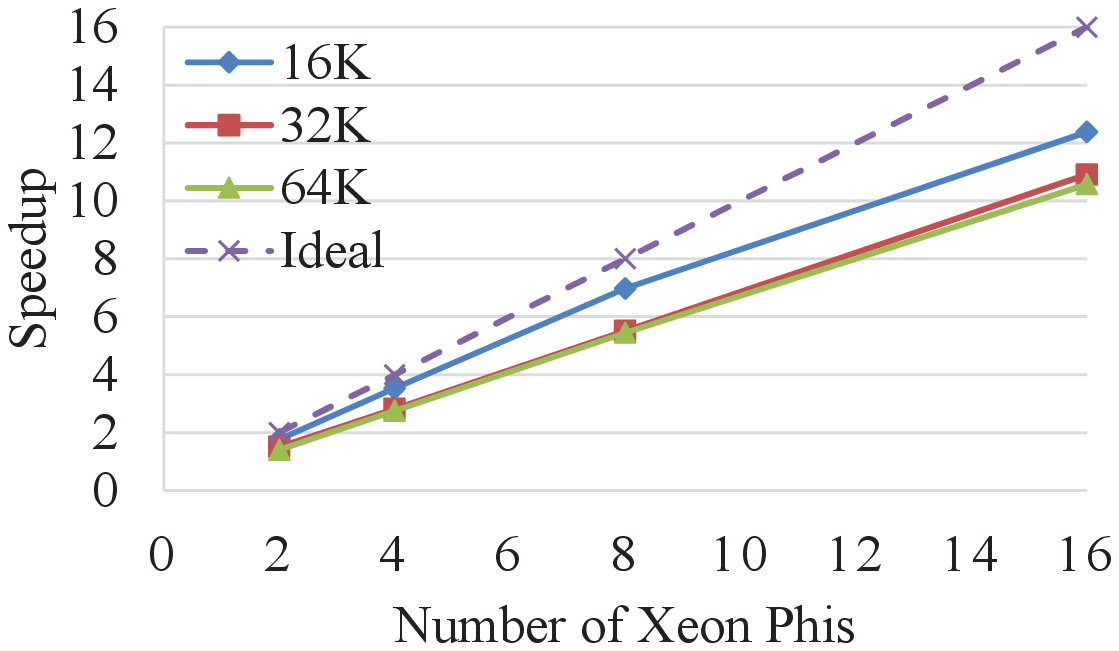}
\caption{}
\label{fig:artificial_scale}
\end{subfigure}
\begin{subfigure}{0.49\linewidth}
\fontsize{7}{7.1}\selectfont
\includegraphics[width=0.985\linewidth]{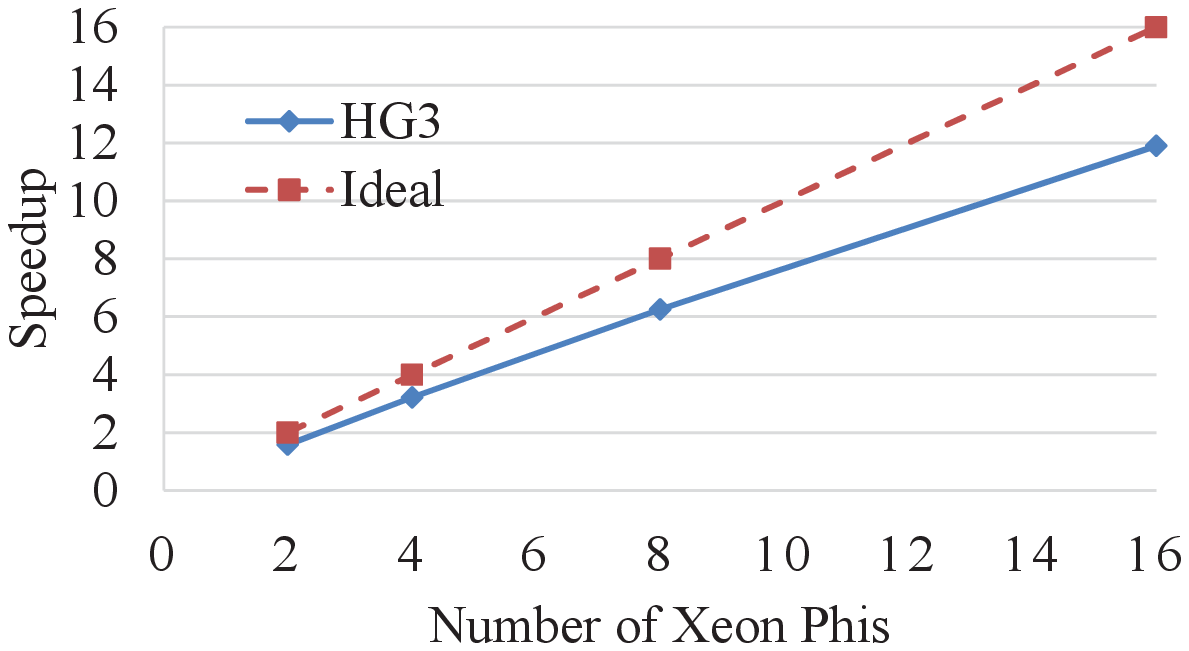}
\caption{}
\label{fig:real}
\end{subfigure}
\caption{Parallel scalability on artificial and real data}
\end{figure}

As for parallel scalibility with respect to varied number of Phis (see Fig. \ref{fig:artificial_scale}), LightPCC achieves an average speedup of 1.6 by using 2 Phis, 3.0 by using 4 Phis, 6.0 by using 8 Phis and 11.3 by using 16 Phis, compared to the execution on a single Phi. Accordingly, the maximum speedup is 1.8, 3.5, 7.0 and 12.4, respectively.
\subsection{Evaluation on Whole Human Genome Expression Data}
We further used a real whole human genome expression dataset to conduct performance comparison. This real dataset is obtained from SEEK \cite{zhu2015targeted}, a computational gene co-expression search engine that supports queries against very large transcriptomic data collections and also publicizes thousands of human datasets from diverse microarray and high-throughput sequencing platforms (available at \url{http://seek.princeton.edu}). This dataset consists of 17,555 genes of 5,072 samples each and is extracted from the GPL570 gene expression data collection produced by Affymetrix Human Genome U133 Plus 2.0 Array.
\begin{table}[!t]
\centering
\caption{Runtimes and speedups on a real human dataset}
\label{tab:human}
\begin{tabular}{|l||l||l||l||l|l|}
\hline
\multirow{2}{*}{LightPCC}& 	\multirow{2}{*}{Time(s)}&	\multicolumn{3}{c|}{Speedup over}\\
\hhline{~~---}
&	&	ALGLIB&	MKL(1 core)&	MKL(16 cores)\\
\hline
1 Phi&		32.1&		13.7&	4.7&	0.4	\\
\hline
2 Phis&		20.3&		21.6&	7.4&	0.6\\
\hline
4 Phi&		10.0&	44.0&	15.0&		1.1\\
\hline
8 Phis&		5.1&	85.4&	29.2&		2.2\\
\hline
16 Phis&	2.7&	162.8&	55.6&		4.3\\
\hline
\end{tabular}
\end{table}

On this real dataset, compared to ALGLIB, LightPCC runs $13.7\times$ faster on a single Phi, $21.6\times$ faster on 2 Phis, $44.0\times$ faster on 4 Phis, $85.4\times$ faster on 8 Phis, and $162.8\times$ faster on 16 Phis (see Table \ref{tab:human}). \hl{In comparison to single-threaded MKL, LightPCC runs $4.7\times$ faster on a single Phi and $55.6\times$ faster on 16 Phis. When it comes to 16-threaded MKL, LightPCC is not able to outperform the former until $\ge 4$ Phis are used, similar to the assessment using artificial datasets. For this case, our algorithm reaches a maximum speedup of 4.3 with 16 Phis. Furthermore, we evaluated MKL on a single Phi as well. The experimental result showed that it took 11.6 seconds for Phi-based MKL to finish the computation, resulting in a speedup of 2.77 over our algorithm on one Phi.} As for parallel scalability, LightPCC also demonstrates good performance (refer to Figure \ref{fig:real}), where compared to the execution on one Phi, the speedup is 1.6 on 2 Phis, 3.2 on 4 Phis, 6.2 on 8 Phis and 11.9 on 16 Phis, respectively.
\section{Conclusion}
PCC is a correlation measure investigating linear relationship between continuous random variables and has been widely used in Bioinformatics. For instance, one popular application is to compute pairwise correlation between gene expression profiles and then build a gene co-expression network to identify common regulation and thus common functions. In addition, PCC can be applied to some computational problems (e.g. feature selection \cite{chandrashekar2014survey} and correlation clustering \cite{pan2015parallel}) in machine learning as well.

In this paper, we have presented LightPCC, the first parallel and distributed all-pairs PCC computation algorithm on Phi clusters. It is written in C++ template classes and harnesses three levels of parallelism (i.e. SIMD-instruction-level parallelism, thread-level parallelism and accelerator-level parallelism) to achieve high performance. Furthermore, we have proposed a general framework for workload balancing in symmetric all-pairs computation. This framework assigns unique identifiers to jobs in the upper triangle of the job matrix and builds bijective functions between job identifier and coordinate space.

\hl{We have evaluated the performance of LightPCC using a set of gene expression profiles and further compared it to a sequential C++ implementation in ALGLIB and an implementation using the {\tt cblas\_dgemm} GEMM routine in MKL, both of which run on the CPU. Performance evaluation showed that LightPCC runs up to $20.6\times$ faster than ALGLIB on one 5110P Phi and up to $218.2\times$ faster on 16 Phis, with a corresponding speedup of up to 6.8 on one Phi and up to 71.4 on 16 Phis over single-threaded MKL. Besides, LightPCC yielded good parallel scalability with respect to varied number of Phis. As part of our future work, we plan to apply this work to construct genome-wide gene expression network (e.g. from conventional microarray data \cite{zhao2015microarray}, emerging RNA-seq data \cite{ballouz2015guidance} \cite{issac2015abstract} or diverse genomic data \cite{wang2014similarity}) and integrate it with statistical and graph analysis methods to identify critical pathways. In addition, our current implementation does not distribute PCC computation onto CPU cores. Therefore, we expect to further boost its performance by employing an alternative CPU-Phi coprocessing model that enables concurrent workload distribution onto both CPUs and Phis.}
%Finally, besides co-expression network, PCC can be used to feature selection based on feature ranking in machine learning \cite{hall1999correlation} \cite{chandrashekar2014survey}. In this context, PCC can be used to measure feature-class or feature-feature relevancy. When two features are perfectly correlated, it suffices by using only one feature to represent the data. In other words, one feature can be deemed as redundant since it does not provide any extra information than the other. By reducing irrelevant and redundant features without incurring much loss of information, we can reduce the computational cost and even improve downstream analysis performance. Nonetheless, note that the computation of feature-class or feature-feature correlation is also not computationally trivial for the cases with larger numbers of features and samples (e.g. text-document classification \cite{forman2003extensive} \cite{baharudin2010review}).
% % use section* for acknowledgement
\section*{Acknowledgment}
This research is supported in part by US National Science Foundation under IIS-1416259 and an Intel Parallel Computing Center award. \textit{Conflict of interest}: none declared.

% Generated by IEEEtran.bst, version: 1.13 (2008/09/30)

\end{document}